\documentstyle[12pt]{article}

\begin{document}





\title{Modified frequentist determination of confidence intervals and its 
equivalence to the 
Bayesian method}

\author{S.I. Bitioukov and N.V. Krasnikov
\\
INR RAS, Moscow  117312}

\maketitle

\begin{abstract}
We propose  modified frequentist definition for the determination of confidence intervals 
for the case of Poisson statistics. Namely, we require that
$\displaystyle 1-\beta^{'} \geq \sum_{n=o}^{n_{obs}+k} P(n|\lambda) \geq \alpha^{'} $. 
We show that this definition is equivalent to
the Bayesian method with prior $\pi(\lambda) \sim \lambda^{k}$. We also 
propose modified frequentist definition for the case of nonzero background. 

\end{abstract}

\newpage

In high energy physics one of the standard problems \cite{1} is the determination of the 
confidence intervals for the parameter 
$\lambda$ in  Poisson distribution
\begin{equation}
P(n|\lambda) = \frac{\lambda^n}{n!}\exp(-\lambda)
\end{equation} 
from the observed number $n_{obs}$ of events. 
There are two methods to solve this problem - the frequentist and the Bayesian.
In Bayesian method  \cite{1,2} due to Bayes theorem 
the probability density for the $\lambda$ parameter 
is determined as
\begin{equation}
p(\lambda |n_{obs}) = \frac{P(n_{obs} |\lambda)\pi(\lambda)}{\int_{0}^{\infty} (P(n_{obs} |\lambda^{'})
\pi(\lambda^{'})d \lambda^{'}} \,.
\end{equation}
Here $\pi(\lambda)$ is the prior function and in general it is not known that is the 
main problem of the Bayesian method. Formula (2) reduces the statistics problem to the 
probability problem. At the  ($1 -\alpha$) probability level the parameters 
$\lambda_{up}$ and $\lambda_{down}$ are determined from the equation \footnote{Usually 
$\alpha$ is taken equal to $0.05$.}
\begin{equation}
\int_{\lambda_{down}}^{\lambda_{up}} p(\lambda|n_{obs}) d \lambda = 1 -\alpha
\end{equation} 
and the unknown  parameter $\lambda$ lies between $\lambda_{down}$ and $\lambda _{up}$ 
with the probability $1 - \alpha$. 
The solution of the equation (4) is not unique. 
One can define 
\begin{equation}
\int_{\lambda_{up}}^{\infty} p(\lambda|n_{obs}) d \lambda = \alpha^{'} \,,
\end{equation}
\begin{equation}
\int_{0}^{\lambda_{down}} p(\lambda|n_{obs}) d \lambda = \beta^{'} \,.
\end{equation}
In general the  parameters $\alpha^{'}$ and $\beta^{'}$ are arbitrary except 
the evident equality 
\begin{equation}
\alpha^{'} + \beta^{'} = \alpha\,.
\end{equation}

The most popular are the following options \cite{1}:

1. $\lambda_{down} = 0$ - upper limit.

2. $\lambda_{up} = \infty$ - lower limit.

3. $\displaystyle \int_{0}^{\lambda_{up}} p(\lambda|n_{obs}) d \lambda = 
\int_{\lambda_{down}}^{\infty} p(\lambda|n_{obs}) d \lambda = \frac{\alpha}{2}$ 
- symmetric interval.

4. The shortest interval -  $p(\lambda|n_{obs})$ inside the interval is bigger or equal to  
$p(\lambda|n_{obs})$ outside the interval.

In frequentist approach the Neyman belt construction \cite{3} is used for the determination 
of the confidence intervals. 
Namely, for each $\lambda$   we require that
\begin{equation}
\sum_{n=n_{-}(\lambda)}^{n_{+}(\lambda)} P(n|\lambda) \geq 1 - \alpha \,.
\end{equation}

For the observed number of events $n_{obs}$ the equation (7) allows to determine the 
confidence interval $[\lambda_{down}(n_{obs}), \lambda_{up}(n_{obs})]$ of possible parameters 
$\lambda$. Note that as in Bayesian approach the choice of $n_{+}(\lambda)$ and  $n_{-}(\lambda)$ 
is not unique. 
In general we have the following equations \cite{4,5,6}
for the determination of $\lambda_{down}$ and $\lambda_{up}$ :
\begin{equation}
\sum_{n=n_{obs}}^{\infty}P(n,\lambda_{down}) = \beta^{'}\,,
\end{equation}
\begin{equation}
\sum_{n=0}^{n_{obs}} P(n,\lambda_{up}) = \alpha^{'}\,,
\end{equation}
where
\begin{equation}
\alpha^{'} + \beta^{'} = \alpha \,.
\end{equation}

In this paper we show that the modified frequentist definition of the confidence interval is 
equivalent to the Bayesian approach. Consider the probability to observe 
the number of events $n \leq n_{obs}$
\begin{equation}
P_{-}(n_{obs}|\lambda) = \sum_{n=0}^{n_{obs}}P(n|\lambda) \,.
\end{equation}
To determine possible values $\lambda_{down}$ and  $\lambda_{up}$ of the confidence interval
we require that
\begin{equation}
1-\beta^{'} \geq P_{-}(n_{obs}|\lambda) \geq \alpha^{'} \,,
\end{equation}
where $\alpha^{'} + \beta^{'} = \alpha$.
The equations for the determination $\lambda_{up}$ and $\lambda_{down }$ have the form
\begin{equation}
P_{-}(n_{obs}|\lambda_{up}) = \alpha^{'}\,,
\end{equation}
\begin{equation}
P_{-}(n_{obs}|\lambda_{down}) = 1 - \beta^{'}\,.
\end{equation}
Note that as in the case of Bayesian approach 
the choice of  $\alpha^{'} $ and $\beta^{'}$ is not unique.
Due to the identity \cite{6}
\begin{equation}
P_{-}(n_{obs}|\lambda) = \int_{\lambda}^{\infty}P(n_{obs}|\lambda^{'})d\lambda^{'}
\end{equation}
the confidence interval $[\lambda_{down}, \lambda_{up}]$ 
is determined from the equations
\begin{equation}
\alpha^{'} = \int_{\lambda_{up}}^{\infty}  P(n_{obs}|\lambda^{'}) d\lambda^{'}\,,
\end{equation}
\begin{equation}
\beta^{'} = \int_{0}^{\lambda_{down}}P(n_{obs}|\lambda^{'})d\lambda^{'}\,.
\end{equation}
The   parameter $\lambda$ lies in the interval
\begin{equation}
\lambda_{down} \leq \lambda \leq \lambda_{up}
\end{equation} 
with the probability $(1- \alpha^{'} - \beta^{'})$.
Due to equations (16,17) our modified frequentist definition (12) is equivalent to Bayes 
definitions (3,4,5) with flat prior $\pi(\lambda) = 1$, namely:
\begin{equation}
\int_{\lambda_{down}}^{\lambda_{up}} P(n_{obs}|\lambda^{'})d\lambda^{'} = 1 - \alpha^{'} - \beta^{'} \,.
\end{equation}

The coverage of the definition (12) means the following. For a hypothetical 
ensemble of similar experiments the probability to observe the number of events 
$n\leq  n_{obs}$ satisfies the inequalities (12). 
 As we noted before the choice of
$\lambda_{down}$ and  $\lambda_{up}$ is not unique. Probably the most natural choice is the use of 
the ordering principle. According to this principle 
 we require  that the probability density $P(n_{obs}|\lambda)$ inside 
the confidence interval $[\lambda_{down}, \lambda_{up}]$ is bigger or equal to 
the probability density outside this interval. For Poisson distribution this requirement 
leads to the formula
\begin{equation}
P(n_{obs}|\lambda_{down}) = P(n_{obs}|\lambda_{up})\,
\end{equation}
for the determination of $\lambda_{up}$ and $\lambda_{down}$.
For such ordering principle $\alpha^{'}$ and $\beta^{'}$ are not independent quantities. It is 
natural to use $\alpha = \alpha^{'} + \beta^{'}$ as a single free parameter. 
Note that the  equations (9) and (13) for the determinations of an upper limit 
$\lambda_{up}$ in frequentist and modified frequentist approach coincide whereas 
the equations (8) and (14) are different. Namely, the equation (14) is equivalent 
to the equation 
\begin{equation}
\sum_{n=n_{obs}+1}^{\infty}P(n|\lambda) = \beta^{'}\,.
\end{equation}
So we see that the summation in our modified equation (21) starts from $n_{obs} + 1$ 
whereas in classical frequentist equation (8) the summation starts from $n_{obs}$.
Classical frequentist equation (8) is equivalent to the Bayes equation (5) with prior 
$\pi(\lambda) \sim \frac{1}{\lambda}$.

 Note that for the case of continuous variable 
$x$ ( $ +\infty > x > -\infty $) our inequality (12) takes the form
\begin{equation}
1 - \beta^{'} \geq \int_{-\infty}^{x_{obs}}f(x^{'}|\lambda)d x^{'} \geq \alpha^{'} \,.
\end{equation} 
Here $f(x,\lambda)$  is the probability 
density function and $ \lambda$ is some unknown parameter. The equations for the determination of 
$\lambda_{up}$ and $\lambda_{down}$ are
\begin{equation}
\int_{-\infty}^{x_{obs}}f(x^{'}|\lambda_{up})dx^{'} = \alpha^{'}\,,
\end{equation}
\begin{equation}
\int_{x_{obs}}^{\infty} f(x^{'}|\lambda_{down})dx^{'} = \beta^{'}\,.
\end{equation}
The equations (23,24) coincide with classical Neyman belt equations.

It is possible to generalize our modified frequentist definition (12), namely:
\begin{equation}
1 - \beta^{'}   \geq P_{-}(n_{obs}|\lambda;k)  \geq \alpha^{'}\,,
\end{equation}
where
\begin{equation}
P_{-}(n_{obs}|\lambda;k) \equiv \sum _{n = 0}^{n_{obs} +k}P(n|\lambda)
\end{equation}
and  $k = 0, {\pm}1, {\pm} 2, ...$

One can find that formulae  (25,26) lead to Bayes equations (4,5) with the 
prior function $\pi(\lambda) \sim \lambda^{k}$.
We can further generalize formulae (25,26) by the introduction 
\begin{equation}
P_{-}(n_{obs}|\lambda; c_{k}) \equiv \sum_{k}c^2_{k}P_{-}(n_{obs}|\lambda;k) \,,
\end{equation}
where $\sum_{k} c^2_{k} = 1$.
 Again we require that
\begin{equation}
1 - \beta^{'} \geq P_{-}(n_{obs}|\lambda ;c_{k}) \geq \alpha^{'}\,.
\end{equation}
One can find that our definition (28) is equivalent to Bayes approach with prior function
\begin{equation}
\pi(\lambda) =\sum_{k} c^2_{k}l_{k} \lambda^{k} \,,
\end{equation}
where
\begin{equation}
l_{k} = \frac{n!}{(n+k)!}\,.
\end{equation}

For the case when we have nonzero background the 
parameter $\lambda$ is represented 
in the form
\begin{equation}
\lambda = b + s \,.
\end{equation}
Here $b \geq 0$   is known background  and $s$ is unknown signal. 
In Bayes approach the generalization of the formula (2) reads
\begin{equation}
p(s |n_{obs},b) = \frac{P(n_{obs} |b+s)\pi(b,s)}{\int_{0}^{\infty} (P(n_{obs} |b+s^{'})
\pi(b,s^{'})d s^{'}} \,.
\end{equation}
For flat prior we have
\begin{equation}
p(s |n_{obs},b) = \frac{P(n_{obs} |b+s)}{\int_{b}^{\infty} (P(n_{obs} |\lambda^{'})
d \lambda^{'}} \,.
\end{equation}
So we see that the main effect of nonzero background is the appearance 
of the factor 
\begin{equation}
K(n_{obs},b) = \int_{b}^{\infty} P(n_{obs} |\lambda^{'})
d \lambda^{'}  
\end{equation}
in the denominator of formula (33). For zero background 
$K(n_{obs}, b =0) = 1$.  One can interpret the appearance of additional factor 
$K(n_{obs},b)$ in terms of conditional probability. Really, for flat prior the 
$P(n_{obs},\lambda)d\lambda $ is the probability that parameter $\lambda$ lies 
in the interval $[\lambda, \lambda +d\lambda]$. For the case of nonzero  background 
$b$ parameter $\lambda = b + s \geq b$.
The probability that $\lambda \geq b$ is equal to $p(\lambda \geq b|n_{obs}) =
K(n_{obs},b)$. The conditional probability that $\lambda$ lies in the interval 
$[\lambda, \lambda +d\lambda]$ provided $\lambda \geq b$ is determined by 
the standard formula of the conditional probability
\begin{equation}
p(\lambda,n_{obs}|\lambda \geq b)d\lambda = 
\frac{p(\lambda,n_{obs})}{p(\lambda \geq b)}d\lambda  = 
\frac{p(\lambda,n_{obs})}{K(n_{obs} s)}d\lambda
\end{equation} 
and it coincides with the Bayes formula (33).

In the frequentist approach the naive  generalization of the inequality (12) is
\begin{equation}
1-\beta^{'} \geq P_{-}(n_{obs}|s +b) \geq \alpha^{'} \,.
\end{equation}
One can show that 
\begin{equation}
1 - \alpha^{'} - \beta^{'} = \int^{b+s_{up}}_{b+s_{down}}P(n_{obs}|\lambda^{'})d\lambda^{'} 
\leq \int^{\infty}_{b}P(n_{obs}|\lambda^{'})d\lambda^{'}\,.
\end{equation}
However the main drawback of the definition (36) is that the probability that the signal $s$ lies 
in the interval $0 \leq s \leq \infty$ is  equal 
to $\int^{\infty}_{  b}P(n_{obs}|\lambda^{'})d\lambda^{'}$ and it 
is less than unity for nonzero background $s>0$ that contradicts to the intuition that 
the full   probability that the signal $s$ lies between
zero and infinity must be equal to unity. To cure this drawback let us require that 
\begin{equation}
1-\beta^{'} \geq 
\frac{P_{-}(n_{obs}|s +b)}{P_{-}(n_{obs}|b)} 
\geq \alpha^{'} \,.
\end{equation}
The inequality (38) leads to the equations for the determination of $s_{down}$ and $s_{up}$ 
which coincide with the corresponding Bayes equations. The generalization of the inequalities 
(38) is straightforward, for instance the inequality (28) reads

\begin{equation}
1 - \beta^{'} \geq 
\frac{P_{-}(n_{obs}|b+s ;c_{k})}{P_{-}(n_{obs}|b ;c_{k})} 
\geq \alpha^{'} \,.
\end{equation}
Upper limit on the signal $s$ derived from the inequality (38) coincides with 
the upper limit in $CL_{s}$ method \cite{7,8}.

Note that frequentist equations (8,9) for $\lambda_{up} = \lambda_{down}$ 
don't satisfy the evident equality 
$\alpha^{'} + \beta^{'} = 1 $.
The natural generalization of the equations (8,9) looks as follows 
\begin{equation}
P_{-1}(n_{obs}|\lambda_{up}) = \alpha^{'} \,,
\end{equation}
\begin{equation}
P_{+1}(n_{obs}|\lambda_{down}) = \beta^{'} \,,
\end{equation}
where
\begin{equation}
P_{-1}(n_{obs}|\lambda) = \sum_{n=0}^{n_{obs}-1}P(n|\lambda) ~+ \frac{1}{2}P(n_{obs}|\lambda)\,,
\end{equation}
\begin{equation}
P_{+1}(n_{obs}|\lambda) = \sum_{n=n_{obs}+1}^{\infty}P(n|\lambda) ~+ \frac{1}{2}P(n_{obs}|\lambda)\,.
\end{equation}
Note that
\begin{equation}
P_{-1}(n_{obs}|\lambda) +  P_{+1}(n_{obs}|\lambda) = 1
\end{equation}
and $\alpha^{'} +\beta^{'} = 1$ for $\lambda_{up} = \lambda_{down}$.
The equations (40,41) are equivalent to the Bayes equations (4,5) with prior 
\begin{equation}
\pi(\lambda) = \frac{1}{2}(1 + \frac{n_{obs}}{\lambda})\,.
\end{equation}
The modified frequentist definition (12) takes  the form
\begin{equation}
1 - \beta^{'} \geq P_{-1}(n_{obs}|\lambda) \geq \alpha^{'} \,.
\end{equation}

To conclude let us stress our main result. For Poisson distribution 
we have proposed modified frequentist definition of the confidence interval and have shown 
the equivalence of the modified frequentist approach and Bayes approach.

 This work has been supported by RFBR grant N 10-02-00468а.

\end{document}